\documentclass[useAMS,usenatbib]{mn2e}
\usepackage[dvips]{graphicx}
\usepackage{color}

\title[On the origin of the scatter around the Fundamental Plane]{On
  the origin of the scatter around the Fundamental Plane: correlations
  with stellar population parameters}
\author[Gargiulo et al.]{A. Gargiulo$^{1}$\thanks{E-mail: gargiulo@na.astro.it}, C. P. Haines$^{2}$, P. Merluzzi$^{3}$, R. J. Smith$^{4}$, F. La Barbera$^{3}$,  \and G. Busarello$^{3}$, J. R. Lucey$^{4}$, A. Mercurio$^{3}$  and
M. Capaccioli$^{5}$\\
$^{1}$Physics Department, Universit\'a ``Federico II'', Napoli, Italy\\
$^{2}$School of Physics and Astronomy, University of Birmingham, Edgbaston, Birmingham, B15 2TT, UK\\
$^{3}$INAF - Osservatorio Astronomico di Capodimonte, via Moiariello, 1-80131 Napoli, Italy\\
$^{4}$Department of Physics, Durham University, Durham DH1 3LE, UK\\
$^{5}$VSTceN, via Moiariello, 1-80131 Napoli, Italy\\
}
\begin{document}

\date{Accepted 2009 March 20. Received 2009 March 20; in original form 2009 January 16}

\pagerange{\pageref{firstpage}--\pageref{lastpage}} \pubyear{2002}

\maketitle

\label{firstpage}

\begin{abstract}
We present a fundamental plane (FP) analysis of 141 early-type
galaxies in the Shapley supercluster at $z{=}0.049$ based on
spectroscopy from the AAOmega spectrograph at the AAT and photometry
from the WFI on the ESO/MPI 2.2m telescope. The key feature of the
survey is its coverage of low-mass galaxies down to
$\sigma{\sim}50$\,km\,s$^{-1}$. We obtain a best-fitting FP relation
$r_{\rm e}{\propto}\sigma_{0}^{1.06\pm0.06}{\langle}I{\rangle}_{\rm
  e}^{-0.82\pm0.02}$ in the $R$ band. The shallow exponent of
$\sigma_{0}$ is a result of the extension of our sample to low
velocity dispersions. Considering the subsample of $ \sigma_{0}>100$
km s$^{-1}$ galaxies, the FP relation is $r_{\rm
  e}{\propto}\sigma^{1.35} {\langle}I{\rangle}_{\rm e}^{-0.81}$,
consistent with previous studies in the high-luminosity regime. We
investigate the origin of the intrinsic FP scatter, using estimates of
age, metallicity and $\alpha$/Fe. We find that the FP residuals
anti-correlate ($>$3$\sigma$) with the mean stellar age in agreement
with previous work. However, a stronger ($>$4$\sigma$) correlation
with $\alpha$/Fe is also found.  These correlations indicate that
galaxies with effective radii smaller than those predicted by the FP
have stellar populations systematically older and with $\alpha$
over-abundances larger than average, for their $\sigma$. Including
$\alpha$/Fe as a fourth parameter in the FP, the total scatter
decreases from 0.088\,dex to 0.075\,dex and the estimated intrinsic
scatter decreases from 0.068 \,dex to 0.049\,dex. Thus, variations in
$\alpha$/Fe account for $\sim$30$\%$ of the total variance around the
FP, and $\sim$50$\%$ of the estimated intrinsic variance.  This result
indicates that the distribution of galaxies around the FP are tightly
related to the enrichment, and hence to the timescale of
star-formation. Our results appear to be consistent with the merger
hypothesis for the formation of ellipticals which predicts that a
significant fraction of the scatter is due to variations in the
importance of dissipation in forming merger remnants of a given mass.
\end{abstract}


\section{Introduction}

Early-type galaxies are observed to obey a set of scaling relations
that connect their photometric and kinematic properties (e.g. Kormendy
relation, Kormendy 1977, Faber-Jackson relation, Faber $\&$ Jackson
1976). Among these, the most notable, due to its surprising small
scatter (${\sim}0.1$\,dex), is the relation between effective radius
$r_{\rm e}$, mean surface brightness within the effective radius
${\langle}I{\rangle}_{\rm e}$ and central velocity dispersion
$\sigma_{0}$ \citep{d87,dressler}. In the three-dimensional space
($\log$$r_{\rm e}$, $\log\sigma_{0}$, $\log{\langle}I{\rangle}_{\rm
  e}$), elliptical galaxies populate a tight plane known as the
fundamental plane (FP) and usually expressed in the form:
\begin{equation}
  \log r_{\rm e} = \alpha \log \sigma_{0} + \beta \log
       {\langle}I{\rangle}_{\rm e} + \gamma.
  \label{formula}
\end{equation}
If elliptical galaxies formed a homologous family, i.e.  systems with
density, luminosity and kinematical structures equal over the entire
early-type sequence and with constant mass-to-light ratios, then the
virial theorem predicts a correlation with $\alpha{=}2$,
$\beta{=}{-}1$. However, observations show that the plane is somewhat
``tilted'' with respect to virial expectations, with best-fit scalings
$\alpha{\sim}1.3$, $\beta{\sim}{-}0.8$ \citep[e.g.][]{j96}.

The origin of the FP tilt has been much debated and can be interpreted
as the breakdown of either of the two assumptions in the virial
expectation. A systematic variation in the mass-to-light ratio along
the FP could be due to variations in the stellar content (age,
metallicity or IMF) and/or the amount of dark matter among
ellipticals. Performing detailed dynamical analyses of 25 galaxies
with SAURON integral-field stellar kinematics to $r_{\rm e}$,
\citet{cappellari} find the ``tilt'' almost exclusively due to real
M/L variations of the form $(M/L){\propto}M^{0.27\pm0.03}$, while
structural and dynamical non-homologies have negligible effects. They
also find the variation of the dynamical M/L ratio to correlate with
the H$\beta$ line-strength, and ascribe most of the tilt to stellar
population (age) effects. On the other hand (e.g. La Barbera et
al. \citeyear{labarbera},Trujillo et al. \citeyear{trujillo}) other
authors find that the tilt is not primarily driven by stellar
populations, but instead results from other effects, such as
non-homology.

Although the FP relation is quite tight, there is none the less a
significant scatter around the plane that cannot be attributed to
measurement errors. The origin of this intrinsic component has
been investigated by many authors. \citet{j96} found that they
were unable to reduce the scatter by introducing additional
parameters, such as ellipticity or isophotal shape of the
galaxies, into the FP relation. Variations in stellar populations
along the sequence of early-type galaxies are found to be
partially responsible for the intrinsic scatter (e.g. Gregg 1992;
Guzman $\&$ Lucey 1993; Guzman, Lucey, Bower 1993). Prugniel and
Simien (1996) studying the correlation between the residuals from
the FP and the residuals from the colour and Mg$_{2}$
line-strength vs. luminosity relations, found that blue and
low-Mg$_{2}$ elliptical galaxies deviate systematically from the
value predicted by the FP. Following this evidence Forbes et al.
(1998), studying a sample of non-cluster galaxies, found that the
residuals of the FP correlate with the ages of the galaxies, i.e.
that the scatter of the FP is partly due to variation in galaxy
age at a given mass, and in particular to variations in the time
of the last starburst. On the contrary, they found that the effect
of changes in metallicity is negligible. Similar results were
obtained by Reda et al. (2005) analysing a sample of isolated
galaxies: some objects deviate from the FP relation having lower
M/L ratio and this was interpreted as due to their younger stellar
populations, probably induced by recent gaseous merger. The same
conclusions were reached by Wuyts et al. (2004) for two
high-redshift clusters.  They found that the residuals from the FP
correlate with the residuals from the H$\beta$ - $\sigma_{0}$
relation. This confirms the role played by stellar populations in
determining the appearance of the FP, with relations appearing
more dispersed for samples of galaxies that are more dispersed in
age.

The existence of the FP, its small observed scatter and the tilt have
presented a long standing challenge to theoretical models explaining
the origin of early-type galaxies.  In fact, whatever the scenario of
formation and evolution of early-type galaxies is, it has to be able
to explain the existence of such a tight correlation and its deviation
from virial expectations and therefore to link galaxy structure and
dynamics with their star-formation histories.

In the recent years, observations and simulations have broadly
supported the galaxy merging scenario which fits naturally into the
$\Lambda$CDM hierarchical cosmology (e.g. Steinmetz \& Navarro,
2002). In the hierarchical scenario ellipticals form through the
merging of disk galaxies \citep{toomre72,toomre77}. In the merging
context, significant new insights have been made through large-scale
gas dynamical simulations of galaxy mergers \citep{robertson},
indicating that for lower mass galaxies, dissipation becomes
increasingly important, driving nuclear starbursts that contribute
larger mass fractions, and producing systematic trends with mass in
both the structures and stellar populations of the remnant ellipticals
\citep{hopkins}. The FP tilt then arises as a direct consequence of
the systematic trends with mass of the importance of dissipation
during mergers.  In the same scenario of galaxy formation, the origin
of the intrinsic scatter in the FP arises as a combination of the
scatter in the total baryon-to-dark matter content of the progenitor
galaxies, and variations in the dissipational fractions at fixed
stellar mass. This latter factor should be observable as correlations
between the residuals from the FP and the stellar population
parameters, and represents a critical test of the merger scenario
\citep{hopkins}, through the predicted coevolution of the stellar
populations and structures of elliptical galaxies.

The increasing importance of dissipation in the formation of low-mass
galaxies and the different mechanisms that drive the evolution and the
star-formation histories for low- and high-mass galaxies (Haines et
al. 2007) should be reflect in variations with mass of the structural
and kinematical properties and hence in variations both in the shape
and orientation of the FP for these two families of galaxies. It
should also be noted that non-merger origins may be important for
lower mass galaxies, whose evolution turns out to be primarly driven
by the mass of their host halo, probably through the combined effects
of tidal forces and ram-pressure stripping (Haines et al. 2007).

\begin{figure*}
  \includegraphics[width=16cm]{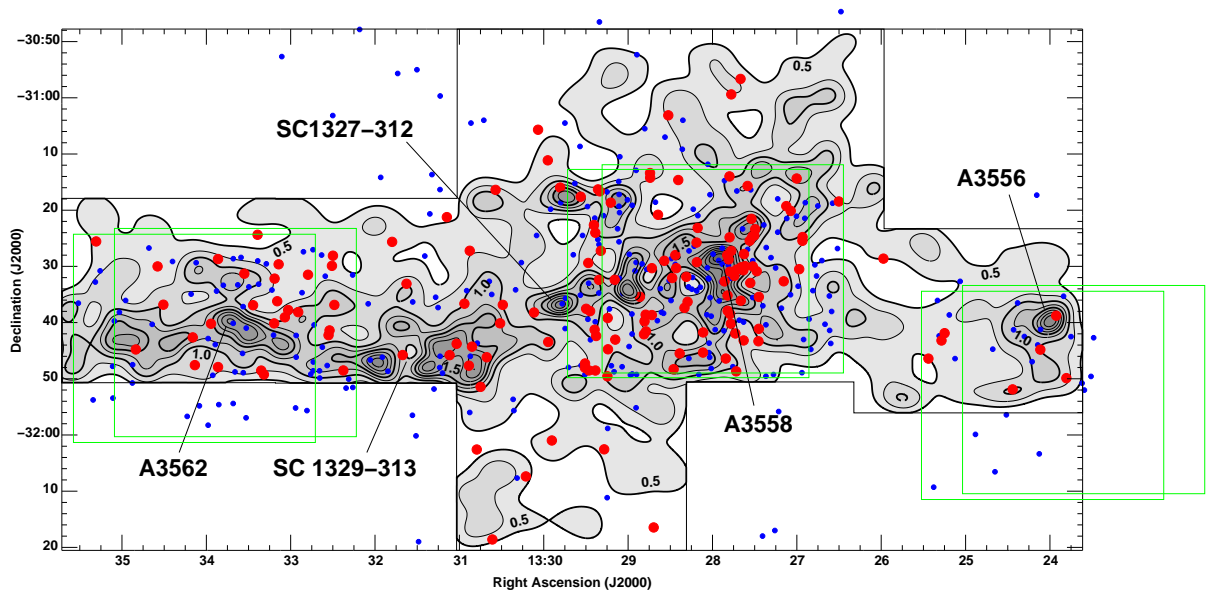}\\
  \caption{Density map of the SOS area (limited by black lines) with
    supercluster galaxies observed spectroscopically but not included
    into the FP sample (blue points) and those that enter the FP (red
    points) indicated. Green boxes show the regions covered by images
    of NFPS. The black contours are isodensity contours of $R{<}21$
    galaxies (see Haines et al. \citeyear{haines06}).}
\label{map}
\end{figure*}

To date no large homogeneous sample of galaxies covering both the
giant and dwarf regime exists.  Although many FP datasets for systems
as different as BCGs, normal Es, dEs, dSphs have been analysed and
compared to look for changes in $\alpha$ \citep{zaritsky}, these
studies suffer from the non-homogenity of the samples both in terms of
differences in measuring the quantities entering the FP (for example
different fits to derive structural parameters, different apertures to
measure the velocity dispersion) as well as the selection of galaxies
themselves. The form of the FP obtained can be influenced by all of
these criteria.

In this paper we present a FP analysis of 141 early-type galaxies from
the Shapley supercluster ($z{\sim}0.049$) with both new $R$-band
surface photometry measurements and published velocity dispersion
measurements from \citet{smith07}.  The sample is randomly selected
down to $M_{R}^{*}+3$ and represents the largest homogeneous sample of
low-mass early-type galaxies with reliable velocity dispersions down
to ${\sim}50$\,km\,s$^{-1}$. In Section 2, we present the photometric
and spectroscopic data (including velocity dispersion measurements and
spectral indices) and the catalogue of ``newly derived'' structural
parameters; the morphological classification is described in Section
3. Our FP fits for the overall and high-$\sigma$
($\sigma{>}100$\,km\,s$^{-1}$) samples are presented in Section 4,
which discusses how the selection criteria can affect the values of
the FP coefficients and mimic a possible curvature of the plane. In
Section 5 we quantify the contribution of stellar population to the
intrinsic scatter. We discuss our results in Section 6 and give a
summary in Section 7.  The origin of the tilt of the FP will be
investigated in a forthcoming paper.

Throughout this paper we use H$_{0}{=}70$\,km\,s$^{-1}$\,Mpc$^{-1}$,
$\Omega_{m}{=}0.3$ and $\Omega_{\Lambda}{=}0.7$. With this cosmology
1 arcsec${=}0.96$\,kpc at $z{=}0.049$ and the distance modulus
is 36.69.

\section{The data}
The sample of galaxies used in this work consists of 141
early-type $R{<}18$ galaxies (red points in Figure~\ref{map})
distributed throughout entire Shapley Optical Survey (SOS,
Mercurio et al. \citeyear{mercurio}) area, but mainly located in
high-density regions (i.e. cluster cores). The sample selection is
summarized in Table 1. In the region covered by SOS there are 378
confirmed supercluster members. Here we analyse the sample of 141
galaxies which have all the following characteristics: (i) an
early-type morphology (see Section 3), (ii) reliable surface
photometry, (iii) a measured velocity dispersion and (iv) an
insignificant H$\alpha$ emission (EW(H$\alpha){<}3$\AA).

\subsection{Spectroscopic data}

Spectra were obtained using the AAOmega fibre-fed spectrograph at
the Anglo-Australian Telescope. A full description is given by
\citet*[hereafter SLH]{smith07}; we summarize some key points
here. The spectroscopic sample limit is $R{=}18$ in the cluster
cores, with galaxies selected from the NOAO Fundamental Plane
Survey \citep[NFPS;][]{smith04} images (green boxes in
Figure~\ref{map}). Outside of these regions, brighter galaxies
were selected from the 2MASS Extended Source Catalogue
($R{\la}15.7$).  The fibre diameter of 2\,arcsec corresponds to
1.9\,kpc. Long integrations resulted in high signal-to-noise
ratios, S/N$\sim$45 per \AA\ for $\sigma{<}100$\,km\,s$^{-1}$ and
S/N$\sim$90 per \AA\ for $\sigma{>}100$\,km\,s$^{-1}$. Velocity
dispersions were measured with respect to the best-matching simple
stellar population (SSP) templates. The errors were estimated from
Monte Carlo simulations, and are ${\sim}0.05$\,dex at
$\sigma{<}100$\,km\,s$^{-1}$ and ${\sim}0.01$\,dex for
$\sigma{>}100$\,km\,s$^{-1}$.  The spectral resolution of
3.2\,\AA\ (82\,km\,s$^{-1}$ instrumental dispersion) allows the
recovery of velocity dispersions as low as $\sim$40\,km\,s$^{-1}$.
However, some galaxies in the sample remain kinematically
unresolved, i.e. have velocity dispersions consistent with zero.
When comparing the new velocity dispersion measurements with those
previously obtained from the NFPS (having a factor ${\sim}3$ lower
S/N), SLH find the new velocity dispersions of low-$\sigma$
objects to be systematically lower by ${\sim}0.10$\,dex, which
they attribute to a combination of higher signal-to-noise and the
use of a range of SSP templates rather than individual K-giant
stars. The absorption index data are tabulated by SLH. Here we
also make use of single-burst equivalent ages, metallicities (Z/H)
and abundance ratios ($\alpha$/Fe) estimated using the models of
Thomas, Maraston $\&$ Bender (2003). Details of this process, and
the stellar population parameters for each galaxy are provided by
Smith, Lucey $\&$ Hudson (2009).
The resulting scaling relations between the three stellar
population parameters and velocity dispersion are:
  \begin{equation}\label{eqage}
   {\rm age}\propto\sigma^{0.43 \pm 0.05}, Z/H \propto \sigma^{0.32
     \pm 0.04}, \alpha/{\rm Fe} \propto \sigma^{0.20 \pm 0.03}.
  \end{equation}
 Typical errors for galaxies with $\sigma$ in the range
 50--100\,km\,s$^{-1}$ are 14$\%$ in age, 0.05\,dex in [Z/H] and
 0.04\,dex in $\alpha$/Fe, while they reduce to half these values for
 galaxies with $\sigma{>}150$\,km\,s$^{-1}$.

In this work we refer to velocity dispersions measured into an
aperture of r$_{\rm e}$/8. We have corrected our velocity dispersions
($\sigma$$_{\rm ap}$) acquired with fibres of 1 arcsec radius (r$_{\rm
  ap}$) to the apertures of r$_{\rm e}$/8 following \cite{j95}:
\begin{equation}
\log \frac{\sigma_{\rm ap}}{\sigma_{r_{\rm e}/8}} = -0.04 \log \frac{r_{\rm ap}}{r_{\rm e}/8}.
\label{correction}
\end{equation}
Hereafter, we adopt the notation $\sigma$$_0$ = $\sigma$$_{r_{\rm
    e}/8}$ for the velocity dispersion corrected to this fiducial
aperture.

Starting from the sample of galaxies observed by SLH we select
those 396 galaxies belonging spectroscopically to the Shapley supercluster
($0.039{<}z{<}0.056$, blue and red points in
Figure~\ref{map}).

\subsection{Imaging data}

For photometry we refer to the SOS survey
\citep{mercurio,haines06}. The SOS is based on data acquired with the
WFI camera (4 $\times$ 2 mosaic of 2k $\times$ 4k CCDs with a pixel
scale of 0.238$^{\prime\prime}$ pixels) mounted on the ESO/MPI 2.2-m
telescope at la Silla observatory. $R$-band imaging was acquired in
good seeing conditions (${\rm FWHM}{\sim}0.7$\,arcsec) for eight
contiguous fields covering a region of 2\,deg$^{2}$ centred on the
Shapley supercluster core (see Figure~\ref{map}).  Five exposures were
obtained for each field giving a total exposure time of 1200\,s for
the mosaic images (240\,s $\times$ 5). The reduction was carried out
with the ALAMBIC pipeline (version 1.0, Vandame 2004) and the
catalogue was produced with the SExtractor package \citep{bertin},
plus a set of procedures designed $ad$ $hoc$ to remove spurious
detections (bad pixels, cosmic rays, etc) and correct the photometry
for blended sources. The survey is complete to $R{=}22$ (${\sim}{\rm
  M}^{\star}{+}7$). For more details see \citet{mercurio}. SOS
photometry is available for 378 supercluster galaxies observed by SLH,
all of which are detected at signal-to-noise levels greater than 100
in each exposure, such that reliable structural parameters can be
derived.

\subsection{Structural parameters}

Structural parameters were derived using the software {\sc 2DPHOT}
described by \citet{2dphot}. This is an automated tool measuring both
integrated and surface photometry of galaxies and is furnished with
several tasks to carry out reliable star-galaxy separation, PSF
modelling and estimation of catalogue completeness. The main steps of
the 2DPHOT algorithm are: i) creation of a clean catalogue of the
input image with SExtractor; ii) estimation of the FWHM and the
definition of ``sure stars''; iii) construction of an accurate PSF
model taking into account both possible spatial variations as well as
deviation of stellar isophotes from circularity; iv) derivation of
structural parameters (effective radius $r_{\rm e}$, mean surface
brightness ${\langle}\mu{\rangle}_{\rm e}$, Sersic index $n$, total
magnitude m$_{\rm tot}$, etc.) by fitting galaxy images with 2D
PSF-convolved Sersic models, as well as the measure of the fit
accuracy ($\chi^{2}_\nu$).

The measurement of structural parameters is strictly dependent both on
the PSF model and on the signal-to-noise ratio. Since the mosaic SOS
images are obtained by stacking jittered images to cover the gap
regions between the eight CCDs, the signal-to-noise ratio is not
constant among the images being lower in the overlapping gap
area. Moreover, in the gap area the PSF turns out to be poorly defined
due to its spatial variations.  We removed the galaxies in these
regions from our sample and performed the surface photometry only for
galaxies in highest S/N regions (224 galaxies), where the PSF is well
defined.  We correct our mean surface brightnesses for cosmological
dimming ($d\langle\mu\rangle_{\rm e}$= 0.208 mag/arcsec$^{2}$),
galactic extinction ($dm$= 0.147 mag, Schlegel et al. 1998) and
k-correction ($dm$ = 0.05, Poggianti 1997) and convert from
mag\,arcsec$^{-2}$ to log\,${\langle}I{\rangle}_{\rm e}$ expressed in
physical units $L_{R\odot}$\,pc$^{-2}$ through
$\log{\langle}I{\rangle}_{\rm e}{=}-0.4({\langle}\mu{\rangle}_{\rm
  e}{-}{\rm M}_{R\odot}{-}5$\,log(206265\,pc/10\,pc)) where
M$_{R\odot}$=4.42 is the solar absolute magnitude (Binney $\&$
Merrifield 1998).

\begin{figure*}
  \includegraphics[width=450pt,angle=0]{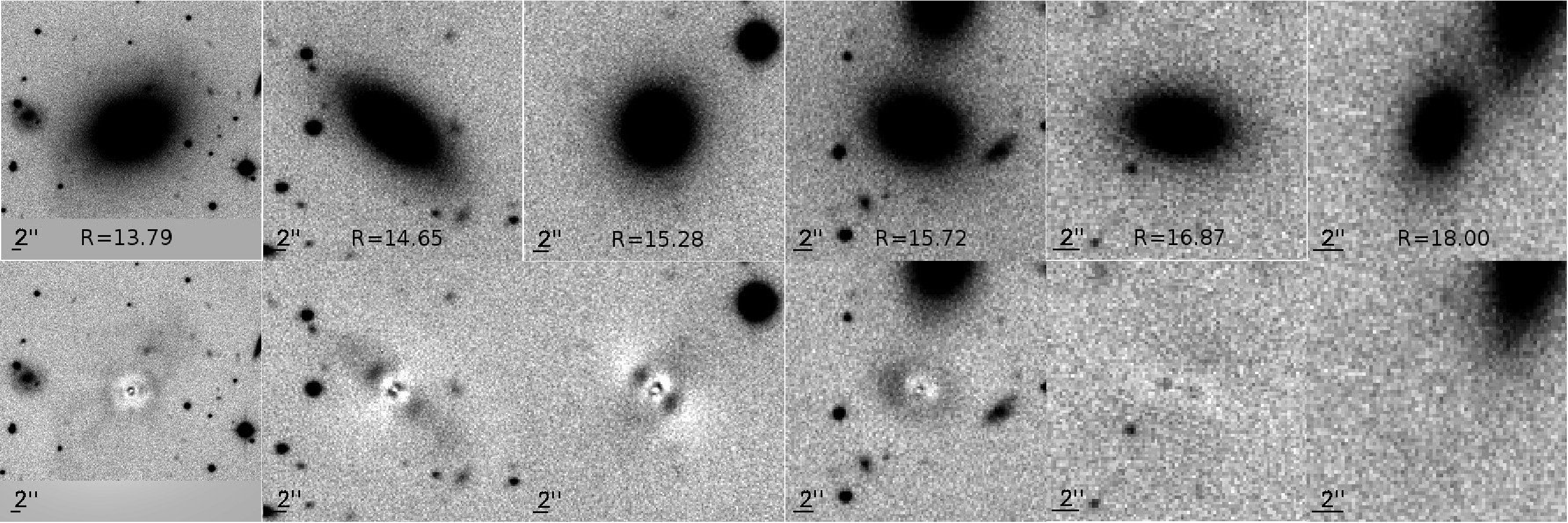}\\
  \vspace{0.5cm}
  \includegraphics[width=450pt,angle=0]{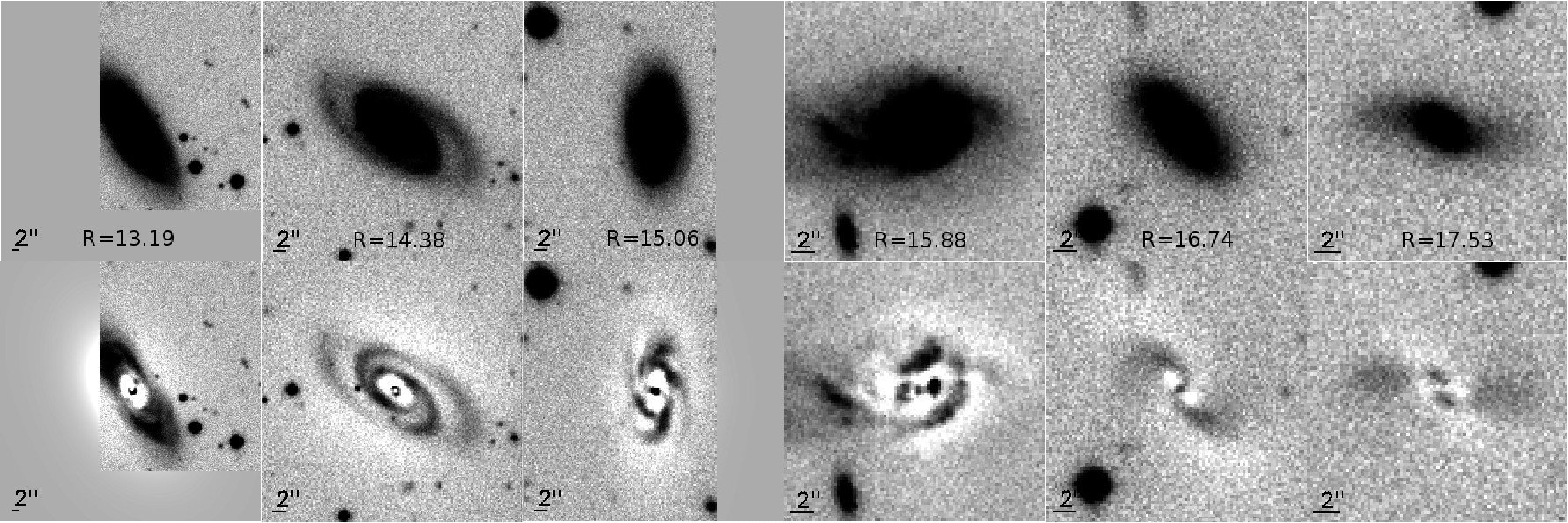}\\
  \caption{{\em Upper panels:} Images of six early-type galaxies
    covering the range of magnitudes of our sample with their
    corresponding residual maps beneath. {\em Lower panels:}
    equivalent for six late-type galaxies. The residual maps
    facilitate detection of spiral arms and other irregularities in
    the late-type ojects.  These examples demonstrate the success of
    2DPHOT in fitting the galaxy profile even when the object is near
    the edge of the image and/or bright objects.}\label{morpho}
\end{figure*}

\subsection{Measuring the uncertainties in effective radius and mean surface brightness}

To derive the errors, we measure the structural parameters on the five
single exposures ($r_{\rm e}^{i}$, $\langle\mu\rangle_{\rm e}^{i}$
with $i{=}1,5$), assuming that the observed spread of their values is
directly related to the S/N on the final mosaic used to derive the
structural parameters entering the FP.

Due to the size of the sample (five exposures) only two scale
estimators can be considered reliable \citep{beers}: the classical
standard deviation and the that obtained by the gapper algorithm
$\sigma_{\rm gap}$. The gapper is a robust scale indicator (Wainer \&
Thissen, 1976) based on the gaps between ordered statistics which has
a high level of efficiency for samples as small as five objects. If we
have $n$ measures of a quantity x ranked in increasing order
(x$_{1}$,x$_{2}$,...x$_{n-1}$,x$_{n}$), according to the gapper
algorithm we can measure a robust scale indicator as:
\begin{equation}
  \sigma_{\rm gap} = \frac{\sqrt{\pi}}{n(n-1)} \sum_{i=1}^{n} w_{i}g_{i},
\end{equation}
 where
\begin{equation}
  g_{i} = x_{i+1}-x_{i}, i=1,...,n-1 \,\,{\rm and}\,\, w_{i} = i(n-i).
\end{equation}
To avoid an overestimation of the errors due to the presence of
outliers, we first compute the $\sigma_{\rm gap}$ of the $\log r_{\rm
  e}^{i}$ and $\log {\langle}I{\rangle}_{\rm e}^{i}$ distributions and
reject all values that deviate more than 3$\sigma_{\rm gap}$ from the
median value, before subsequently computing the classical standard
deviation as well as the covariance matrices of the clipped sample for
both variables ($\sigma_{\log r_{\rm e}^{i}}$, $\sigma_{\log
  {\langle}I{\rangle}_{\rm e}^{i}}$, cov[$\log r_{\rm e};\log
  {\langle}I{\rangle}_{\rm e}$]).

The errors on the mosaic value of $\log r_{\rm e}$ and $\log
{\langle}I{\rangle}_{\rm e}$ ($\delta{\log r_{\rm e}}$ and
$\delta_{\log {\langle}I{\rangle}_{\rm e}}$ respectively) are given by
$\sigma_{\log r_{\rm e}^{i}}/\sqrt{n}$ and $\sigma_{\log
  {\langle}I{\rangle}_{\rm e}^{i}}/\sqrt{n}$, where $n$ is the number
of measures available.  The typical errors on the logarithms of
effective radius ($\log r_{\rm e}$) and mean surface brightness ($\log
{\langle}I{\rangle}_{\rm e}$) are 0.03 and 0.04, respectively. These
errors explicitly include the effects of noise in the galaxy, but not
the presence of neighbouring objects.  We have estimated the effect of
the latter, by repeatedly placing copies of the galaxies one-by-one at
random positions across the same CCD image (where the PSF should
remain constant) and reapplying 2DPHOT, finding the variations in the
structural parameters to be consistent with the previously obtained
errors, albeit with a small number (${\sim}2$\%) of ${>}5{\sigma}$
outliers when the galaxy is placed very close to a bright star or
galaxy.

\section{Morphological classification}

We have morphologically classified the 224 galaxies with available
surface photometry, by inspection of the residual maps provided by
2DPHOT. We denote as ``late-type" all galaxies showing signs of spiral
arms or asymmetric disturbance, and as ``early-type" all those with no
such structures.  The resolution of SOS images does not allow any
finer classification, for example into elliptical and lenticular
galaxies, since the presence of a residual disk can be seen only in
particular cases, i.e. when it is very bright, widespread or
edge-on. In Figure \ref{morpho} we show some illustrative examples of
galaxies classified as early- and late-types, covering the full range
of magnitudes studied here. We have checked this classification by
comparing our results with those of \citet{thomas06} for a subsample
of 54 galaxies belonging to A3558 and A3562, finding perfect
agreement.  In their classification galaxies were subdivided into E,
S0, Se and Sl classes (the latter being early and late spirals). Our
early-type sample contain only galaxies classified as E and S0 by
\citet{thomas06}.

\begin{figure}
\begin{center}
  \includegraphics[width=160pt,angle=-90]{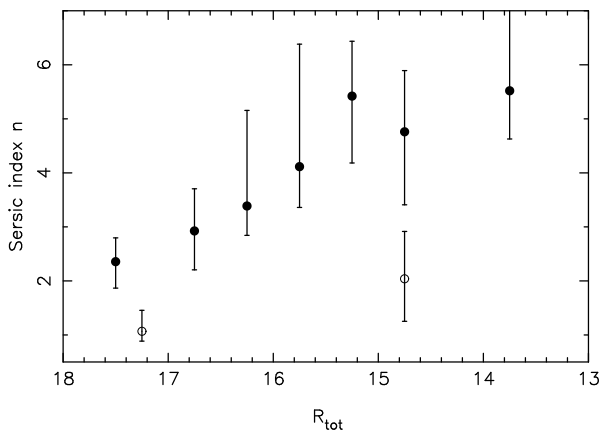}\\
  \caption{Distribution of Sersic index as function of morphological
    type and magnitude. The filled circles indicate the median value of
    $n$ for early-type galaxies in each magnitude bin (chosen to
    contain the same number of galaxies), while the ``error bars"
    indicate the interquartile values. The open circles correspond
    instead to late-type galaxies.}\label{n_dist}
\end{center}
\end{figure}

The strong correlation between the Sersic index $n$ and luminosity
\citep{young,caon} prevents any morphological classification based on
Sersic index alone. \citet{graham} show that the values of Sersic
index $n$ of dE, ordinary E/S0 galaxies and BCGs follow a $continuous$
trend from $n{<}0.5$ to $n{\sim}10$, whereby brighter galaxies have
larger values of $n$. In Figure~\ref{n_dist} we demonstrate the
equivalent trend for our galaxy sample,
plotting the Sersic index distribution as function of morphological
type and magnitude. The median value of $n$ for early-type (filled
circles) and late-type (open circles) galaxies are shown for each
magnitude bin (chosen to contain the same number of galaxies), while
the ``error bars" indicate the inter-quartile values. As pointed out by
\citet{graham} the distribution of $n$ for early-type galaxies is
shifted towards higher values with increasing luminosity. Late-type
galaxies have lower values of $n$ at all magnitudes, but there is not
a clean \textit{a priori} separation between the two morphological
classes.

The morphological classification results in a sample of 141 early-type
galaxies and 44 late-type or irregular galaxies, while 31 galaxies are
removed from the sample since they are not kinematically resolved, and
8 are removed as having significant H$\alpha$ emission
(EW(H$\alpha){>}3$\AA). We summarize the characteristics of the sample
in Table \ref{sample}, while in Table \ref{sample2} we report
structural and kinematical parameters with the corresponding errors
for the 141 galaxies used in the following section.

\begin{table}
\begin{center}
\caption{Spectroscopic sample: the first four lines indicate the
  criteria used to select the galaxy sample, and the number of
  galaxies remaining after each step. In the follwoing lines, the 224
  remaining galaxies are then divided into four classes, of which only
  the 141 early-type objects are used in our FP
  analysis.} \label{sample}
\bigskip
\begin{tabular}{c c}
\hline \hline \noalign{\smallskip}
Sample selection\\
\noalign{\smallskip}
\hline \hline \noalign{\smallskip}
Galaxies with spectra & 565\\
\noalign{\smallskip} \hline \noalign{\smallskip}
Cluster members ($0.039{<}z{<}0.055$) & 396\\
\noalign{\smallskip} \hline \noalign{\smallskip}
Cluster members in SOS region& 378\\
\noalign{\smallskip} \hline \noalign{\smallskip}
Cluster members in SOS region with surface photometry & 224 \\
\noalign{\smallskip} \hline \hline \noalign{\smallskip}
Galaxy classification\\
\noalign{\smallskip} \hline \hline \noalign{\smallskip}
early-type & 141 \\
late-type & 44 \\
\noalign{\smallskip} \hline \noalign{\smallskip}
unresolved kinematically & 31 \\
\noalign{\smallskip} with EW(H$\alpha){>}3$$\AA$ & 8\\
\noalign{\smallskip} \hline \hline
\end{tabular}
\end{center}
\end{table}

\begin{table*}
\caption{Structural and kinematical parameters for the FP galaxy
  sample.{\em Column 1}: ID (SLH); {\em Columns 2 and 3}: RA(J2000)
  and DEC(J2000); {\em Column 4}: total $R$-band magnitude; {\em
    Columns 5 and 6}: log $\sigma_{0}$ referred to an aperture of
  r$_{\rm e}$/8 radius and $\delta_{\log\sigma_{0}}$; {\em Columns 7
    and 8}: log r$_{\rm e}$ and $\delta_{\log r_{\rm e}}$; {\em
    Columns 9 and 10}: log ${\langle}I{\rangle}_{\rm e}$ and
  $\delta_{\log {\langle}I{\rangle}_{\rm e}}$; {\em Column 11}: $n$;
  {\em Column 12}: reduced $\chi$$^{2}_\nu$. The whole catalogue will
  be available in the electronic version.}
\bigskip \label{sample2} {\scriptsize
\begin{tabular}{lccccccccccc} \hline \hline \noalign{\smallskip} ID & RA(J2000) & Dec(J2000) & R$_{\rm tot}$ &\ \ $\log\sigma_{0}$\ \ &\ \ $\delta_{\log\sigma_{0}}$\ \ & $\log r_{\rm e}$ & \ \ $\delta_{\log r_{\rm e}}$\ \ & \ \ \ \ $\log
  {\langle}I{\rangle}_{\rm e}$\ \ \ \  &\ \ $\delta_{\log I_{\rm
      e}}$\ \ & $n$ & $\chi^{2}$ \\ & & & & (km/s) & & (kpc) & &
  \ \ \ \ $L_{\odot}$\,pc$^{-2}$\ \ \ \ & & & \\ \noalign{\smallskip}
  \hline \hline \noalign{\smallskip}
TMASSJ13274662-3059237 & 13:27:46.6 & -30:59:24 & 15.166 & 2.201 & 0.010 & 0.904 & 0.042 &
  1.766 & 0.070 & 10.708 & 1.25 \\
NFPJ132828.6-313205 &  13:28:28.6 & -31:32:04 & 15.404 & 2.112 & 0.012 & 0.308 & 0.002 & 2.861 & 0.004 & 1.589 & 6.43\\
NFPJ132738.0-313041 & 13:27:38.0 & -31:30:41 & 16.549 & 2.044 & 0.016 & 0.038 & 0.004 & 2.943 & 0.008 & 2.283 & 1.31\\
NFPJ133408.2-314735 & 13:34:08.1 & -31:47:34 & 16.083 & 1.757 & 0.097 & 0.682 & 0.044 & 1.842 & 0.059 & 3.518 & 2.03\\
TMASSJ13295423-3201001 &  13:29:54.2 & -32:00:59 & 15.229 & 2.153 & 0.010 & 0.475 & 0.005 & 2.597 & 0.007 & 4.617 & 1.17\\
NFPJ132656.0-312528 & 13:26:56.0 & -31:25:27 & 14.813 & 2.031 & 0.007 & 0.797 & 0.007 & 2.118 & 0.011 & 5.452 & 1.99\\
TMASSJ13294834-3115580 & 13:29:48.3 & -31:15:58 & 14.075 & 2.296 & 0.004 & 0.984 & 0.023 & 2.043 & 0.035 & 6.945 & 1.15\\
NFPJ132810.5-312310 & 13:28:10.5 & -31:23:09  & 13.868 & 2.267 & 0.004 & 1.108 & 0.043 &  1.875 & 0.067 & 7.796 & 1.48\\
NFPJ132426.5-315153 & 13:24:26.5  & -31:51:53 & 14.857 & 2.302 & 0.006 & 0.506 & 0.036 & 2.685 & 0.051 &  8.414 & 1.38\\
NFPJ132923.8-314832 & 13:29:23.2  & -31:48:32 & 15.060 & 1.718 & 0.062 & 0.972 & 0.046 & 1.673 & 0.073 & 9.545 & 1.77\\
\noalign{\smallskip} \hline \hline
\end{tabular}}
\end{table*}

\section{The fundamental plane of R$<$18 Shapley galaxies}

For the FP we use the representation of Equation \ref{formula} where
$r_{\rm e}$ is the effective radius measured in kpc, $\sigma_{0}$ is
the central velocity dispersion corrected to an aperture of radius r$_{\rm
  e}$/8 in km\,s$^{-1}$ (see Section 2.1) and
${\langle}I{\rangle}_{\rm e}$ is the mean surface brightness within
$r_{\rm e}$ expressed in $L_{\odot}$\,pc$^{-2}$.  To derive the value
of $\alpha$, $\beta$ and $\gamma$ we adopted the orthogonal fit which
minimizes the quantity:
\begin{equation}
 \sum_{i=1}^{N} \frac{|\log r_{\rm e}^{i} - (\alpha \log
   \sigma_{0}^{i} + \beta \log {\langle}I{\rangle}_{\rm e}^{i} +
   \gamma)|}{\sqrt{1 + \alpha^{2} + \beta ^{2}}},
\end{equation}
i.e. the sum of the absolute residuals perpendicular to the plane.
This is less sensitive to outliers than the classic least-squares
method \citep*[hereafter JFK96]{j96}. For the full sample of 141
galaxies we found:
\begin{equation}
    \log r_{\rm e}{=}1.03{\pm}0.06\log\sigma_{0}{-}0.82{\pm}0.02 \log
    {\langle}I{\rangle}_{\rm e}{+}0.33,
\end{equation}
where the errors on coefficient are computed via a bootstrap
procedure. The above relation presents a scatter of 0.088\,dex in
the log r$_{\rm e}$ direction.  We estimate the intrinsic
contribution to the scatter by subtracting in quadrature from the
rms scatter, the observational errors on each variable taking into
account the correlation between the errors on the effective radius
and surface brightness. We obtain an intrinsic scatter
$\sigma_{\rm  int}$ = 0.068. When deriving the $\sigma_{\rm int}$
the errors on distance estimated through the standard deviation of
the redshift distribution are also considered.

To take into account the different uncertainties on $\sigma_{0}$,
$r_{\rm e}$, ${\langle}I{\rangle}_{\rm e}$ along the plane, a weighted
fit is adopted. Following \citet{cappellari} we add in quadrature to
the measurement errors the intrinsic scatter orthogonal to the FP,
$\delta_{\rm int\bot}$.
The fitting function to be minimized is:
\begin{equation}
\label {fp_weighted} \sum_{i=1}^{N} w_{i} |\log
  r_{\rm e}^{i} - \alpha \log \sigma_{0}^{i} - \beta \log
  {\langle}I{\rangle}_{\rm e}^{i} - \gamma|,
\end{equation}
where the sum is extended over all sample galaxies and where
\begin{equation}
 w_{i}^{-2}{=}\delta_{\rm int\bot}^{2}{+}\delta_{\log
r_{\rm e}}^{2}\ {+}
(\alpha\delta_{\log\sigma_{0}})^{2}{+}(\beta\delta_{\log
I_{\rm e}})^{2} {-}2\beta{\rm cov}[\log r_{\rm e};\log {\langle}I{\rangle}_{\rm e}].
\end{equation}
The intrinsic scatter, $\delta_{\rm int\bot}$, is adjusted to give the
expected value of 0.8 per degree of freedom (the mean absolute value
of a standardised gaussian distribution is 0.8).  The resulting
equation for the FP is:
\begin{equation}
    \log r_{\rm e}{=}1.06{\pm}0.06\log\sigma_{0}{-}0.82{\pm}0.02 \log
    {\langle}I{\rangle}_{\rm e}{+}0.28.
\end{equation}
\begin{figure*}
  \includegraphics[width=160pt,angle=-90]{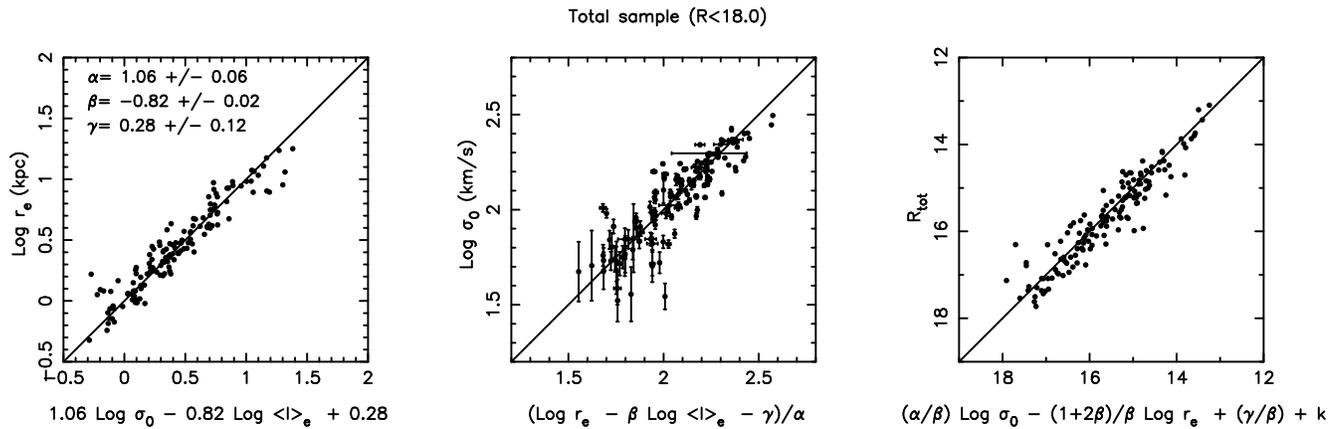}\\
  \caption{{\em Left,} and {\em central panels:} The edge-on views of
    the Shapley $R{<}18$ FP. {\em Right panel}: The edge-on view of
    the FP as it appears along the direction of luminosity. Black
    lines are the best-fit relations. $\alpha$, $\beta$ and $\gamma$
    values are reported in the left panel.}\label{fp_tot}
\end{figure*}
\begin{figure*}
  \includegraphics[width=160pt,angle=-90]{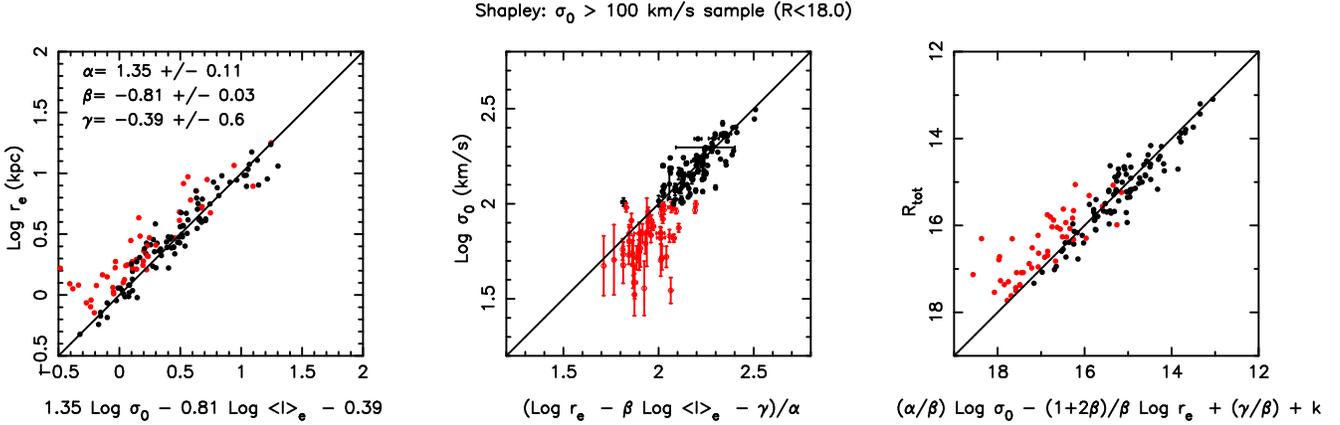}\\
  \caption{{\em Left,} and {\em central panels:} The edge-on views of
    the Shapley $R{<}18$, $\sigma_{0}{>}100$\,km\,s$^{-1}$ (black
    dots) FP. {\em Right panel:} The edge-on view of the FP as it
    appears along the direction of luminosity. Black lines are the
    best-fit line of the high $\sigma$ sample. Galaxies with
    $\sigma_{0}{<}100$\,km\,s$^{-1}$ (red dots) are reported for
    comparison. $\alpha$, $\beta$ and $\gamma$ values are reported in
    the left panel.}\label{fp_high}
\end{figure*}
The orthogonal and weighted fits are consistent, demonstrating that
the results are robust to the effects of the larger $\sigma_{0}$
uncertainties (0.01 dex for $\sigma_{0}>100$ km s$^{-1}$ and 0.05 for
$\sigma_{0}<100$ km s$^{-1}$) for the lowest mass galaxies. The
scatter in the log r$_{\rm e}$ direction is now equal to 0.092, while
its intrinsic component is 0.070. Figure \ref{fp_tot} shows the two
edge-on FP views (left and central panels) and its trend along the
direction of luminosity (right panel).

The value of the $\alpha$ coefficient obtained for the Shapley sample,
is significantly lower than the typical values of 1.2--1.3 reported
for samples dominated by giant galaxies (e.g. JFK96). If we restrict
our sample to galaxies with $\sigma_{0}>$ 100\,km\,s$^{-1}$
(high-$\sigma_{0}$ sample, 91 galaxies) we obtain the FP relation:
\begin{equation}
    \log r_{\rm e}{=}1.35{\pm}0.11\log\sigma_{0}{-}0.81{\pm}0.03\log
    {\langle}I{\rangle}_{\rm e}{-}0.40,
\end{equation}
with the overall scatter in $\log r_{\rm e}$ reduced to 0.090. In
Figure \ref{fp_high} we plot the FP as fitted for the
high-$\sigma_{0}$ sample (black points). Galaxies with
$\sigma_{0}{<}100$ km\,s$^{-1}$ are also reported for comparison (red
dots). The $\alpha$ value of the high-$\sigma_{0}$ sample is closer to
those of other authors. The lower $\alpha$ value found for the total
sample is thus probably due to the extension of our sample to very low
mass galaxies, down to $\sigma_{0}\sim$\,50\,km\,s$^{-1}$. The
improved method used to obtain the velocity dispersions, as well as
the high signal-to-noise levels of the spectra, can produce systematic
effects on the resultant values of $\sigma_{0}$, particularly for
low-$\sigma_{0}$ objects.  SLH find that velocity dispersions obtained
using single, old, solar-metallicity models instead of templates with
different metallicities (as used in this paper) are overestimated by
${\sim}6$\% for $\sigma_{0}{=}75$\,km\,s$^{-1}$ galaxies and by
${\sim}18$\% for $\sigma_{0}{=}50$\,km\,s$^{-1}$. As a result, in our
sample the low-$\sigma_{0}$ values are systematically lower than those
of previous samples (e.g. NFPS sample), causing an increase of the FP
tilt of about 10--15$\%$. The impact of the low-$\sigma_{0}$ limits is
explored further in Section 4.2. We notice that results presented in
Section 5 and discussed in Section 6 has been checked to be
independent of $\sigma_{0}$ cuts.

\subsection{Comparison with Coma}

\begin{figure}
\begin{center}
  \includegraphics[width=7.0cm,angle=-90]{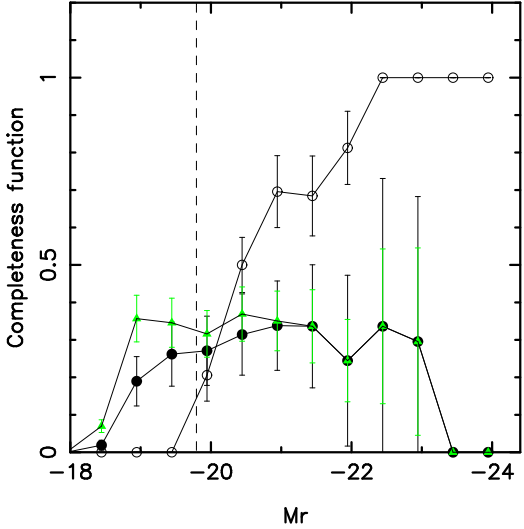}\\
  \caption{The spectroscopic completeness function both for Coma (open
    circles) and for Shapley (filled circles) sample. Green triangles
    show the completeness for all Shapley galaxies observed
    spectroscopically, including those with no successful measurement
    of $\sigma$.} \label{fc_coma}
\end{center}
\end{figure}

\begin{figure}
\begin{center}
  \includegraphics[width=7cm,angle=270]{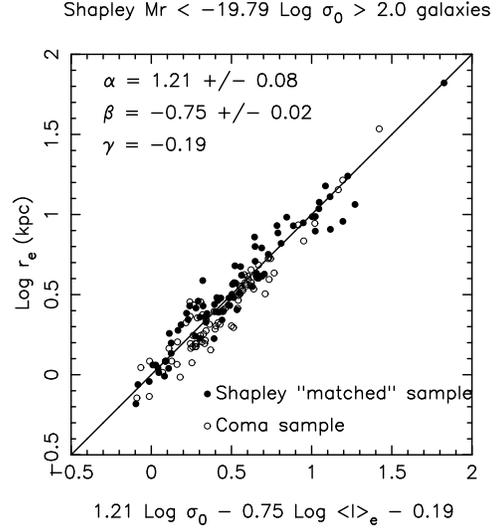}\\
  \caption{Edge on view of the FP for Shapley early-type
galaxies with M$_r{<}{-}19.79$ and
$\sigma_{0}{>}100$\,km\,s$^{-1}$ (Shapley ``matched'' sample,
black points). The solid line represents the Shapley FP fit. For
    comparison are reported also Coma galaxies from JFK96 (open
    points).}\label{fpcoma}
\end{center}
\end{figure}

Besides the range of velocity dispersions, the determination of the
three FP coefficients is strictly dependent on other factors such as
the selection criteria of the sample, the fit algorithm and the
procedure used to derive structural and kinematical parameters
\citep{kelson}. We address this problem comparing our FP with the
previous work by JFK96.  Their sample consists of 81 early-type
galaxies down to Gunn $r{<}15.3$ in the central region of Coma
cluster.  For this section, we have corrected our velocity dispersions
to their standard fixed aperture of 0.595\,$h^{-1}$\,kpc of radius. To
analyse the consistency of the two samples, in Figure~\ref{fc_coma}
we plot the spectroscopic completeness functions (CFs) of the Coma
(open circles) and Shapley (filled circles) FP samples as a
function of magnitude.  The magnitudes of the Shapley galaxies are
converted from $R$ to the Gunn-$r$ photometric system according to
the typical colour of $r{-}R{=}0.35$ for elliptical galaxies
\citep{fukugita}. The Coma CF is computed using the JFK96 FP
sample and the complete photometric catalogue of early-type Coma
galaxies published by \citet{j94}. The limiting magnitude of the
Coma sample is M$_r{=}{-}19.79$ (dashed line). The Coma sample is
complete at bright magnitudes, but the completeness declines
rapidly towards zero at the faint magnitude limit.  In contrast,
about 30\% of Shapley galaxies are spectroscopically observed and
have available velocity dispersion measurements independent of
magnitude down to M$_r{=}{-}19.79$.  In the magnitude bins fainter
than M$_{r}=-20$, the Shapley CF declines gently due to an
increasing fraction of galaxies with surface brightnesses and
velocity dispersions too low for successful $\sigma_{0}$
measurements, before dropping rapidly in the faintest bin. Green
triangles represent the CF of Shapley galaxies observed
spectroscopically with no regards to successful velocity
dispersion measurements: the decline at faintest magnitude has now
disappeared except for the faintest bin (M$_r{>}{-}18.5$). The
distribution reflects well the random selection criteria of our
spectroscopic survey for $R{<}18$ galaxies. To compare the FPs of
the two samples we have selected from Shapley only early-type
galaxies with M$_r{<}{-}19.79$ and
$\sigma_{0}{>}100$\,km\,s$^{-1}$ (hereafter ``matched'' sample)
corresponding roughly to the limits of the Coma FP sample. This
new sample consists of 88 galaxies. The FP of the ``matched''
sample is:
\begin{equation}
    \log r_{\rm e} = 1.21{\pm}0.08 \log \sigma_{0} - 0.75{\pm}0.02 \log
         {\langle}I{\rangle}_{\rm e} -0.19,
\end{equation}
with a scatter equal to 0.08\,dex in log r$_{\rm e}$ direction. In
Figure~\ref{fpcoma}
we plot the edge-on view of the FP for the ``matched'' sample
(filled circles).  Coma galaxies (open circles) are reported for
comparison. Since in JFK96 the value of H$_{0}$ is set equal to
50\,km\,s$^{-1}$\,Mpc$^{-1}$, the effective radii of Coma galaxies
have been shifted by a factor --0.146\,dex. Both the distributions
and the dispersions of the two samples are consistent, in fact the
FP for Coma sample, as found by JFK96 is:
\begin{equation}
    \log r_{\rm e}{=}1.31{\pm}0.07\log \sigma_{0}{-}0.84{\pm}0.02 \log
    {\langle}I{\rangle}_{\rm e}{-}0.082.
\end{equation}
The possible reasons for the remaining slight differences between the two
FPs could be the different sampling at the brightest magnitudes of
the two datasets, and the different procedures used to obtain the
structural parameters, since JFK96 fit their early-type galaxies
with de Vaucouleurs profiles.

\subsection{A curved surface or a selection effect?}

Observing the high-$\sigma_{0}$ fit (see Figure~\ref{fp_high}) it
is notable that all galaxies with velocity dispersions less than
100\,km\,s$^{-1}$ (low-mass galaxies) are systematically displaced
above (left panel) or below (central panel) the best-fit plane as
defined by the high-$\sigma_{0}$ sample. In particular, the
log\,$\sigma_{0}$ edge-on view suggests that low-mass galaxies do
not follow the same relation as massive galaxies. Thus, at face
value, the FP appears curved, as suggested also by other studies
(JFK96, Zaritsky et al. \citeyear{zaritsky}, D'Onofrio 2008,
Nigoche-Nietro 2008) investigating the curvature of the FP as a
function of mass and/or luminosity. Studying a sample of 69 faint
early-type galaxies in the core of Coma cluster, \citet{matkovic}
find a similar behaviour for the $L-\sigma$ relation, with dwarf
galaxies having a trend ($L{\,\propto\,}\sigma^{2.01\pm0.36}$)
shallower than that observed for elliptical giant systems
($L{\,\propto\,}\sigma^{4}$) .

\begin{figure}
\begin{center}
  \includegraphics[width=170pt,angle=-90]{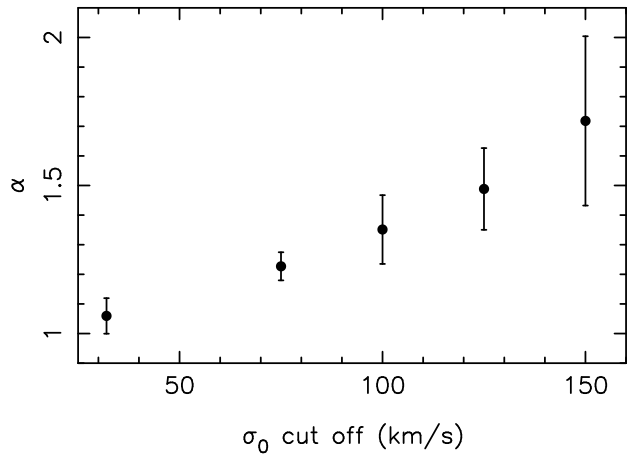}\\
  \caption{Values of the $\alpha$ coefficient for Shapley sample of
    galaxies with increasing lower cuts on velocity dispersion.  The point
    with $\sigma_{0}{<}50$\,km\,s$^{-1}$ represents the coefficient
    for the whole sample.}\label{coeff_alpha}
\end{center}
\end{figure}

\begin{figure}
\begin{center}
  \includegraphics[width=230pt, angle=-90]{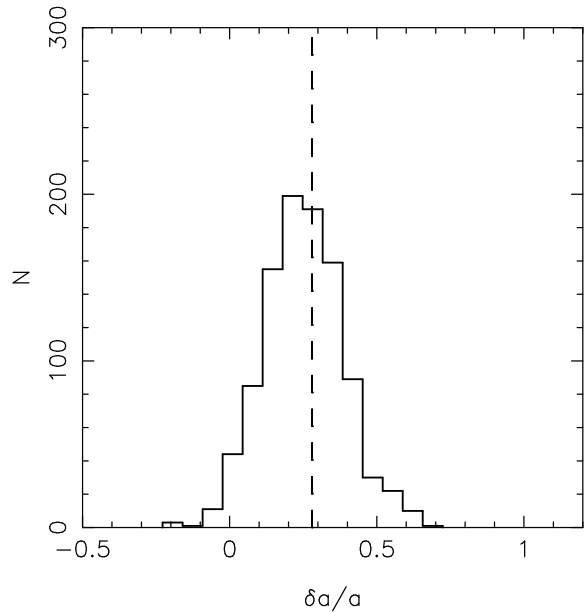}\\
  \caption{The distribution of the relative variation of $\alpha$
    coefficient after applying the $\sigma_0>100$\,km\,s$^{-1}$ cut
    for the simulated sample. The dashed line marks the variation of
    $\alpha$ when applying the $\sigma_{0}$ cut to the Shapley
    data.}\label{sim}
\end{center}
\end{figure}

\begin{figure}
\begin{center}
  \includegraphics[width=230pt,angle=-90]{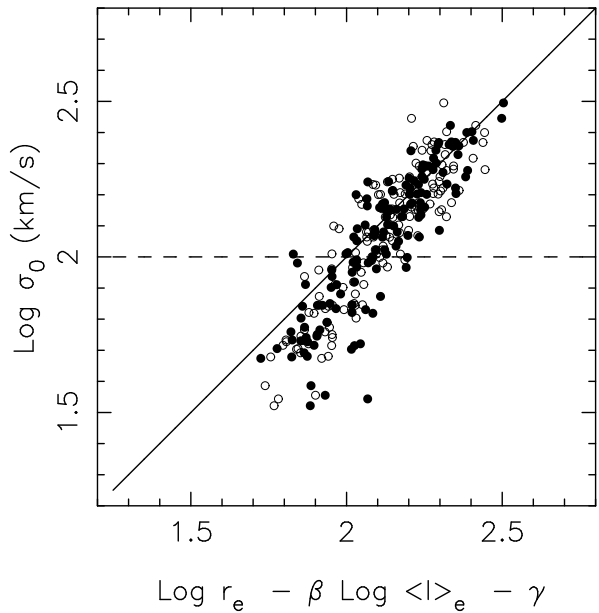}\\
  \caption{The edge-on projection of the FP for the real Shapley
    sample (black dots) and for one of the simulated catalogues
    (open circles). The line represents the best-fitting relation for
    the subset with $\sigma_0>100$\,km\,s$^{-1}$.}
\label{conf_sim}
\end{center}
\end{figure}

In Figure~\ref{coeff_alpha}, we investigate this possible curvature by
analysing the dependence of $\alpha$ as a function of different
low-$\sigma_0$ limits.
We cut the galaxy sample at various values of $\sigma_{0}$ and
estimate the FP coefficients through the weighted orthogonal fit
procedure. One can see that $\alpha$ increases systematically as the
$\sigma_{0}$ cut moves to higher values, removing more galaxies from
the sample. The values of $\alpha$ for the whole sample
($\alpha$=1.06$\pm$0.06) and for the highest-$\sigma_{0}$ cuts
($\alpha$=1.72$\pm$0.28) differ by more than twice the standard
errors, indicating a possible curvature of FP. To investigate if these
observed variations of $\alpha$ with the sample selection are due to a
real curvature of the FP, or induced by selection effects, a set of
simulations were performed. We constructed 1000 mock catalogues using
the observed values of log\,${\langle}I{\rangle}_{\rm e}$ and
log\,$\sigma_{0}$ and the value of log\,$r_{\rm e}$ assigned using the
relation:
\begin{equation}
\label{simulazione} \log r_{\rm e} = \alpha \log \sigma_{0} + \beta
\log {\langle}I{\rangle}_{\rm e} + \gamma + N(0,\delta),
\end{equation}
where, in this case, $\alpha$ = 0.92, $\beta$ = --0.78, and $\gamma$ =
0.496 are the coefficients obtained by fitting the FP of the overall
Shapley sample with log $r_{\rm e}$ as the dependent variable,
$\delta$ = 0.099 is the observed scatter in the log $r_{\rm e}$
direction, and N(0,$\delta$) is a Gaussian random variable with zero
mean and standard deviation of $\delta$. For each mock catalogue we
evaluate the relative change in $\alpha$ ($\delta\alpha/\alpha$)
between the orthogonal fits obtained for the whole sample and after
applying a cut at $\sigma_{0}=100$\,km\,s$^{-1}$.  The resulting
histogram of $\delta\alpha/\alpha$ values is shown in
Figure~\ref{sim},
where the dashed line indicates the $\delta\alpha/\alpha$ computed
directly for the Shapley sample. This shows that the observed
change in $\alpha$ is fully consistent with that expected for a
linear relation. Moreover, in Figure~\ref{conf_sim}
we compare the FP edge-on projection (Eq.~11) for the observed
high-$\sigma_{0}$ sample (solid symbols) and a simulated sample
obeying the overall Shapley FP relation (open symbols). The two
samples show similar behaviour, with the same apparent curvature
such that galaxies with low velocity dispersions are placed
systematically below the FP relation. Given that the simulated
sample is defined to follow a linear relation, this suggests that
the apparent curvature in Figure~\ref{fp_high} can be explained
solely by selection biases. Hence, in spite of the wide-range in
velocity dispersions covered by our sample, this is not sufficient
to distinguish between either a linear or curved FP relation, but
there is no convincing evidence for the latter in our data.

\section{FP residuals and stellar populations}

We find that for the Shapley FP, the contribution of the intrinsic
scatter (0.070) to the total scatter (0.092) is larger than that
of the measurement errors (0.060). Taking advantage of the wide
variety of indices derived by SLH we investigate the possible
origin of the intrinsic scatter of the FP analysing the
correlations between both spectral indices (Mgb, Fe5015, Fe4383,
H$\beta$, HgF, HdF) and stellar population parameters (age,
metallicity, $\alpha$-enhancement) and the residuals from the FP.

\subsection{FP residuals vs. single spectral indices}

\begin{figure*}
  \includegraphics[width=17cm,angle=0]{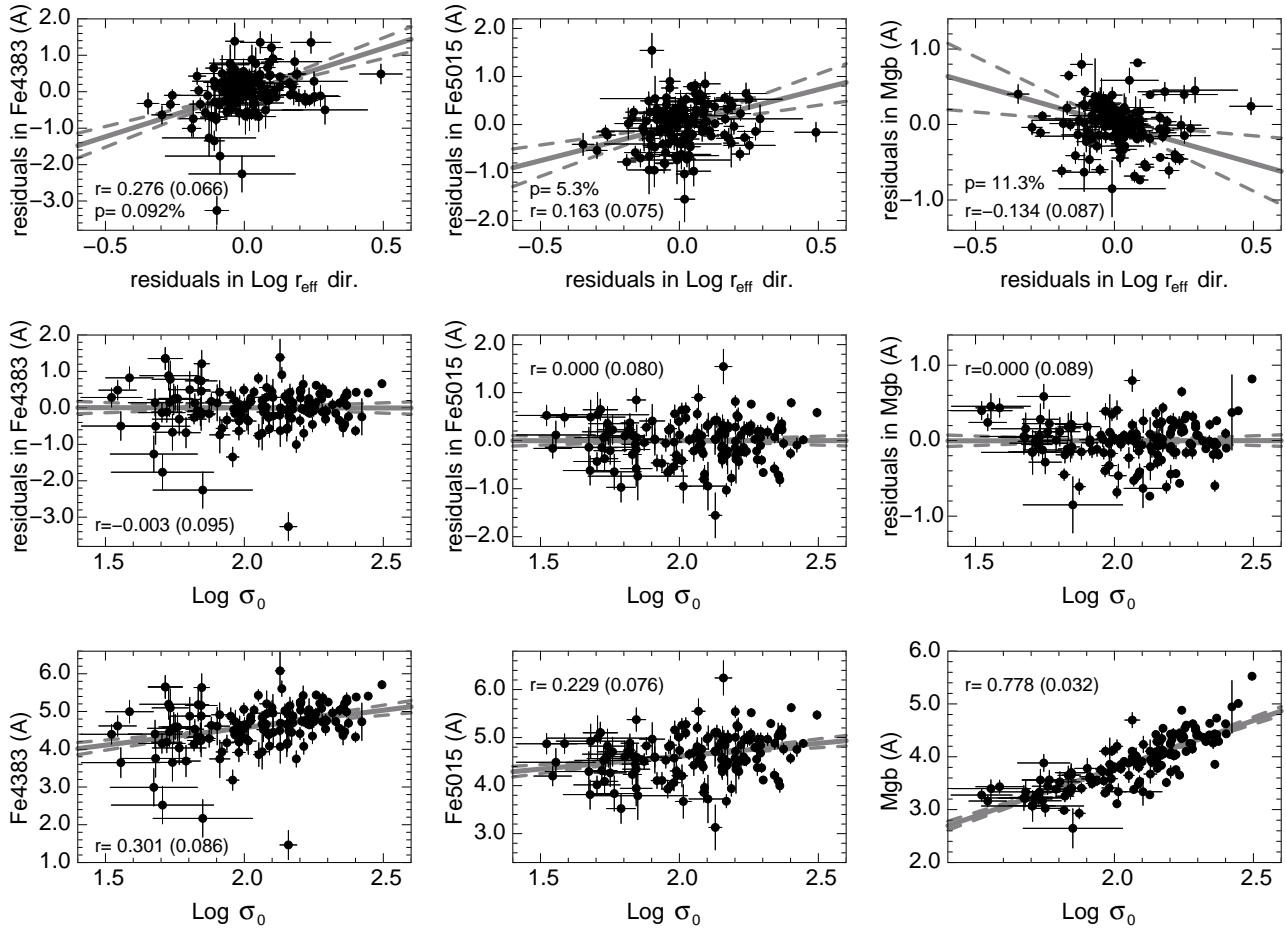}\\
  \caption{{\em Low-left panel}: The Fe 4383-log\,$\sigma_{0}$ relation. In all the panels, the solid (dashed) lines indicate the
mean (and $1\sigma$ confidence limits) bisector least-squared fits
after 10000 Monte Carlo realizations. The corresponding
correlation coefficient $r$ (and uncertainty) is indicated. {\em
Middle-left panel}: The corresponding Fe 4383-log\,$\sigma_{0}$
residuals in Fe 4383 direction respect to the fit (solid line in
low-left panel) plotted against log\,$\sigma_{0}$. {\em Top-left
panel}: Correlations between the residuals of the Fe
4383-log\,$\sigma_{0}$ relation in Fe 4383 direction with the
residuals from the best-fitting FP in the log r$_{e}$ direction.
The probability p that two quantities with correlation factor $r$
are not correlated is also reported. {\em Central panels}: The
same for Fe 5015. {\em Right panels}: The same for Mgb.}
\label{corrindfe}
\end{figure*}

\begin{figure*}
  \includegraphics[width=17cm,angle=0]{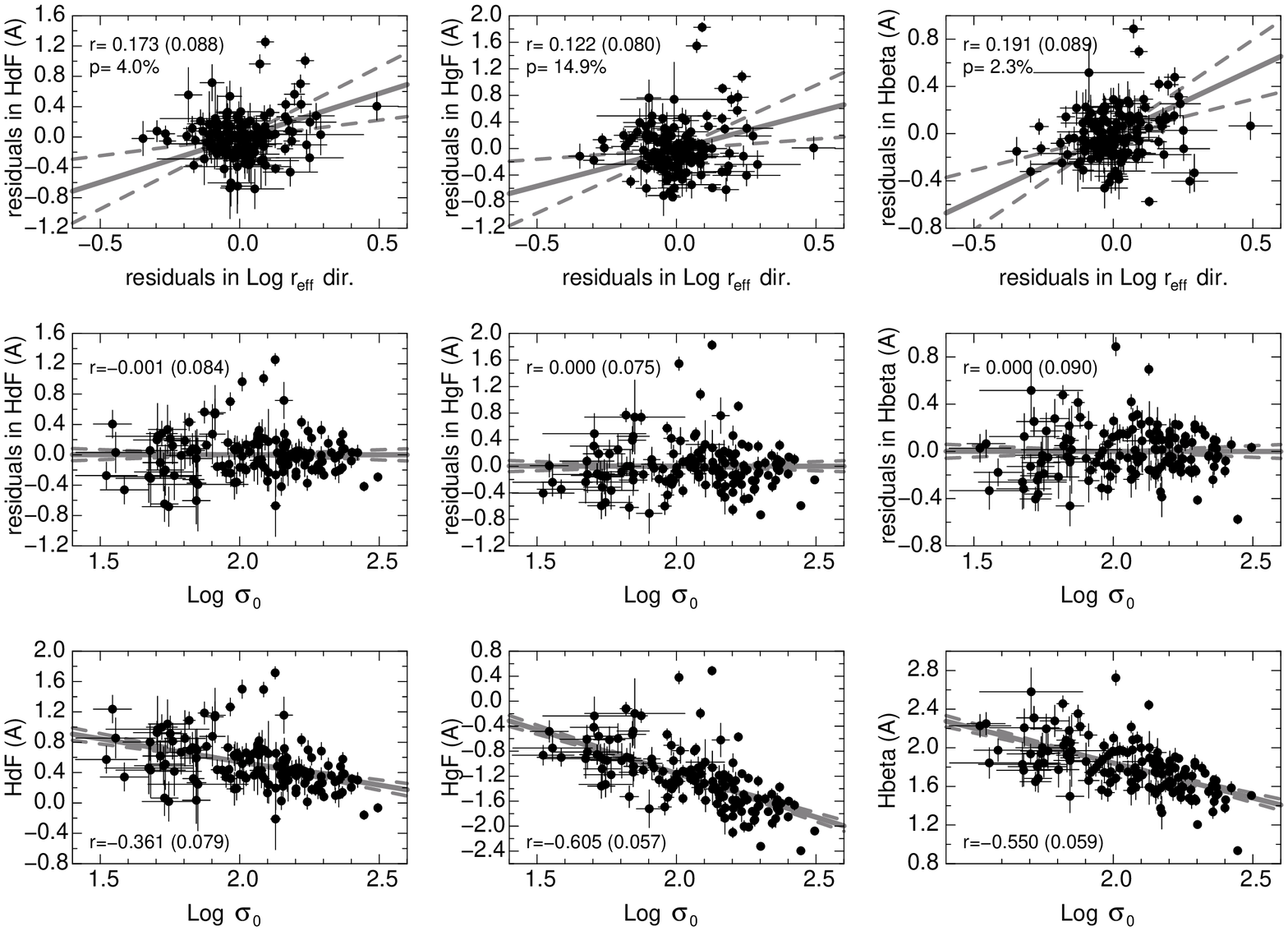}\\
  \caption{As for Figure 11, but now showing correlations for the
    Balmer line indices HdF (left panels), HgF (central panels) and H$\beta$
    (right panels).}
\label{corrindba}
\end{figure*}

All of the many spectral indices are known to correlate with
$\sigma_{0}$, which, if not corrected for, could produce spurious
correlations with the FP residuals. To avoid this problem, rather
than use the spectral indices themselves, we consider their
residuals with respect to the \textit{index}-log $\sigma_{0}$
relation. For each index, we fitted first the \textit{index}-log
$\sigma_{0}$ relation, shown in Figures~\ref{corrindfe} and
~\ref{corrindba} (bottom panels) for metallicity- and age-
sensitive indices respectively, and then determine the residuals
with respect to that relation (middle panels).  In the figures,
the upper panels show the \textit{index}-log $\sigma_{0}$
residuals versus the residuals from the FP in the log r$_{\rm e}$
direction. For each relation, we quantify the product-moment
correlations between the residuals and spectral index through the
correlation factor $r$ and the bisector least-squares fit,
assuming that the distributions are both Gaussian. Uncertainties
in $r$ and the fits are estimated through 10000 Monte Carlo
realizations, taking into account the errors on each value. These
values together with the probability $p$ that two quantities with
correlation factor $r$ are not correlated are reported in
Figures~\ref{corrindfe} and ~\ref{corrindba}.

Among the primarily metal-sensitive indices (Figure 11), the strongest
dependence is found for Fe4383, which is correlated with FP residuals
at the ~4$\sigma$ level. A similar but weaker correlation is obtained
for Fe5015. For Mgb, however, a weak correlation is obtained in the
opposite sense. Thus galaxies which are more compact than expected
from the FP have lower Fe4383 indices than expected for their
$\sigma_{0}$, but {\it higher} Mgb indices than expected for their
$\sigma_{0}$. This already provides a hint that the physical driver
for the residual trends is not the total metallicity, but instead the
ratio between Mg and Fe abundances.  The Balmer lines show weak
positive correlations, such that galaxies with strong index values
(for their $\sigma_{0}$) are more diffuse than predicted by the FP.
This could be caused either by age or by metallicity effects, since
the Balmer lines are sensitive to both parameters to some extent.

\subsection{FP residuals vs. stellar population parameters}

Using a single spectral index it is not possible to disentangle
between the effects of age and metallicity. In this section, we
use estimates of age, metallicity and $\alpha$-enhancement
($\alpha$/Fe) derived from the index measurements by means of a
multi-index procedure, to provide more physically meaningful
information (Smith et al. 2009).

\begin{figure*}
  \includegraphics[width=17cm,angle=0]{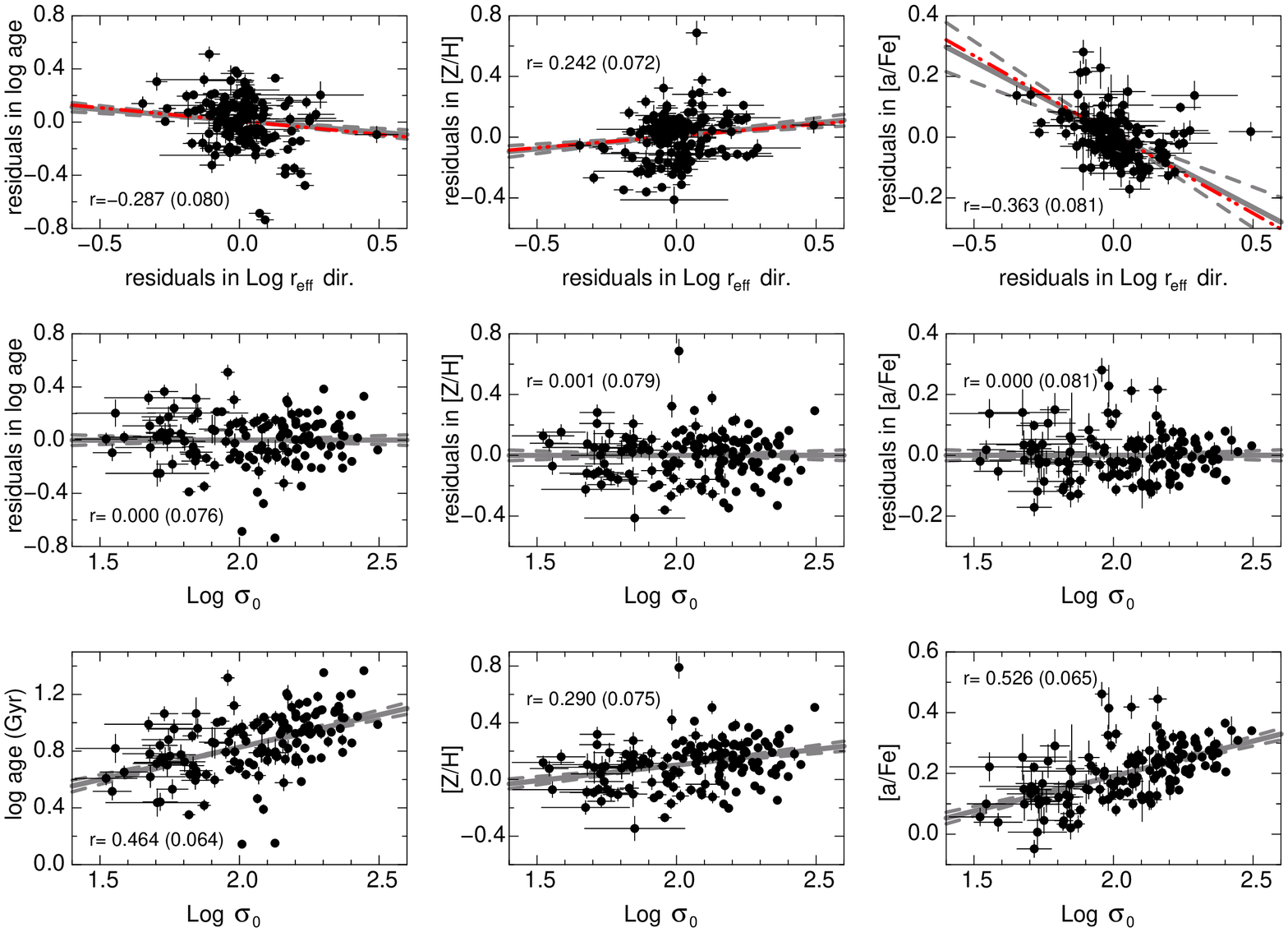}\\
  \caption{As for Figure 11, but now showing correlations for the SSP
    parameters, age (left panels), total metallicity (central panels) and $\alpha$/Fe
    (right panels). The red dot-dashed line shows the relation
    obtained taking into account the correlated errors (see text).} \label{correlations}
\end{figure*}

As for the single spectral indices, the three stellar population
parameters age, metallicity and $\alpha$/Fe correlate strongly with
$\sigma_{0}$ as shown by the scaling relations of Equation
\ref{eqage}, therefore we first compute the residuals of stellar
population parameters with respect to these relations.  The lower
panels of Figure~\ref{correlations} show the stellar population
residuals against $\log \sigma_{0}$ characterized, by construction, by
the absence of correlations between the residuals and log $\sigma_{0}$
itself. The top panels show the resultant correlations between the
age, metallicity and $\alpha$/Fe residuals (left, middle and right
panel respectively) and the FP residuals along the log $r_{\rm e}$
projections.

Both age and $\alpha$/Fe are seen to be strongly anti-correlated (at
$>$3$\sigma$ and $>$4$\sigma$ respectively) with the residuals
computed in the $\log r_{\rm e}$ direction, while metallicity shows a
{\em positive} correlation with the FP residuals along the same axis.
We find these correlations to be independent of the range of $\sigma$
covered, for example obtaining consistent results when using only the
high-$\sigma_{0}$ sample. In Table \ref{percentuale} we summarize the
probability $p$ for all the cases.
\begin{table*}
\begin{center}
\caption{Probability that two quantities with correlation factor
$r$ as in Figure 11, 12 and 13, are not correlated.}
\label{percentuale}
\bigskip
{\scriptsize\begin{tabular}{c c c c c c c c c} \hline \hline
  \noalign{\smallskip} Mgb & Fe5015 & Fe4383 & Hbeta & HgF & HdF & Log
  age & Z/H & $\alpha$/Fe\\ \noalign{\smallskip} \hline
  \noalign{\smallskip} 11.3\% & 5.3\% & 0.092\% & 2.3\% & 14.9\% &
  4.0\% & 0.056\% & 0.38\% & \textbf{0.001\%} \\ \noalign{\smallskip}
  \hline \hline
\end{tabular}}
\end{center}
\end{table*}

We note that in the figures 11, 12 and 13, the velocity dispersion
enters into both axes, since each quantity is a residual from a
relation involving $\sigma_{0}$. The errors are hence correlated
which will produce a bias towards positive correlations. For each
of the primary SSP parameters we measure the level of this bias
through Monte Carlo simulations, based on fake data in which there
are no intrinsic correlations. Firstly, for each galaxy we assign
fake SSP values (age, Z/H, $\alpha$/Fe) following the previously
derived index-$\sigma_{0}$ relations, and having the same rms
intrinsic scatter in the SSP value about the relation. We then
perturb each of the values log $\sigma_{0}$, log $r_{\rm e}$, and
log ${\langle}I_{\rm e}{\rangle}$ by their uncertainties for each
galaxy, and recalculate the residuals about the FP and the
index-$\sigma_{0}$ relation as before. Through these simulations,
we measure this bias ($\Delta$ r), i.e. the difference in the
correlation coefficient, due to the correlated measurement errors
in $\sigma_{0}$ and finds:
\begin{equation}
\Delta r ({\rm log\,age}) = 0.03, \,\, \Delta r ({\rm Z/H}) =
0.02, \,\, \Delta r ({\rm \alpha/Fe}) = 0.04.
\end{equation}
We see that this bias contributes to the observed correlation
between the metallicty and FP residuals, but acts to oppose the
anti-correlations seen for age and $\alpha$-enhancement. In each
case the effect of the bias is small, being at a level of
$\sim$10\% of the observed correlation. The relation between
residuals obtained taking into account this bias is shown in
figure 13 by the red dot-dashed line.

It should be remembered that the stellar population parameters
presented here are derived through fixed apertures covering only the
galaxy centres. Therefore, if galaxies have significant population
gradients, the observed trend could be due to sampling the central
stellar populations at different radii. In the case of metallicity,
this may be a factor, as early-type galaxies are observed to have
negative metallicity gradients (Kuntschner et al. 2006; Rawle et
al. 2008) of the order $-0.20\pm0.05$\,dex in [Z/H]. However,
spatially-resolved spectroscopy of early-type galaxies show that they
generally have flat radial trends in age and $\alpha$/Fe
(S\'anchez-Bl\'azquez et al. 2007; Rawle et al. 2008), indicating that
the anti-correlation of age and abundance ratios with the FP residuals
in the log $r_{\rm e}$ and log ${\langle}I{\rangle}_{\rm e}$
directions is robust.

The strong correlations found between the stellar population
parameters, principally the $\alpha$/Fe, and the residuals of the FP
suggest that its scatter is in part due to variations in the stellar
populations at fixed galaxy $\sigma_0$.  To this aim we fit the
modified FP relation:
\begin{equation}
  \log r_{\rm e} = \alpha \log \sigma_{0} + \beta \log {\langle}I{\rangle}_{\rm e} +
  \kappa {spp_i} + \gamma,
\label{addssp}
\end{equation}
where the stellar population parameter spp$_{i}$ is, in turn, $\alpha$
enhancement, age and metallicity of the galaxies, and study the
variation induced in the scatter by keeping $\alpha$ and $\beta$ fixed
as in Eq.  \ref{simulazione} and allowing $\kappa$ to vary to minimize
the FP scatter.  It is necessary to keep $\alpha$ and $\beta$ fixed as
each of the stellar population parameters correlate strongly with
$\sigma_{0}$, and hence would introduce spurious variations in
$\alpha$. The results obtained are listed in Table \ref{fit_4dim},
where the second column indicates the estimated strength of the spp
term $\kappa$, the following two columns indicate the overall scatter
in the log $r_{\rm e}$ direction after the addition of the spp term,
and the relative intrinsic scatter, and the final column the rms
contribution of the ssp term to the scatter in log $r_{\rm e}$.
\begin{table*}
\caption{The impact on the overall and intrinsic FP scatter of adding
  a further term (ssp) to the orthogonal fit of the fundamental
  plane. $Column$ $1$: The $\kappa$ value which minimizes the overall
  FP scatter along the log $r_{\rm e}$ direction to Eq.~\ref{addssp};
  $Column$ $2$: total scatters around the FP; $Column$ $3$: estimated
  intrinsic FP scatter; $Column$ $4$ estimated rms contribution of the
  spp term in the log $r_{\rm e}$ direction.}
  \bigskip
\begin{tabular}{ccccc} \hline \hline \noalign{\smallskip}
spp & $\kappa$ & rms in log r$_{\rm e}$ dir. & intrinsic FP rms & spp
rms contribution\\ \noalign{\smallskip} \hline \noalign{\smallskip} -
& & 0.088 & 0.068 & \\ \noalign{\smallskip} \hline
\noalign{\smallskip} $\alpha$/Fe & $+0.582\pm0.128$ & 0.075 & 0.049 &
0.047\\ \noalign{\smallskip} \hline \noalign{\smallskip} Log age &
$+0.169\pm0.057$ & 0.084 & 0.063 & 0.026\\ \noalign{\smallskip} \hline
\noalign{\smallskip} Z/H & $-0.160\pm0.061$ & 0.087 & 0.066 &
0.016\\ \noalign{\smallskip} \hline \hline \label{fit_4dim}
\end{tabular}
\end{table*}
The fit with the most significant $\kappa$ coefficient ($4.5\sigma$)
is the fit with the $\alpha$/Fe as fourth parameter. In the case of
age and metallicity, the new coefficient is non-zero at just the
$\sim$3$\sigma$ level, indicating the weaker impact of these
parameters on the fundamental relation.

\begin{figure*}
  \includegraphics[width=6cm,angle=-90]{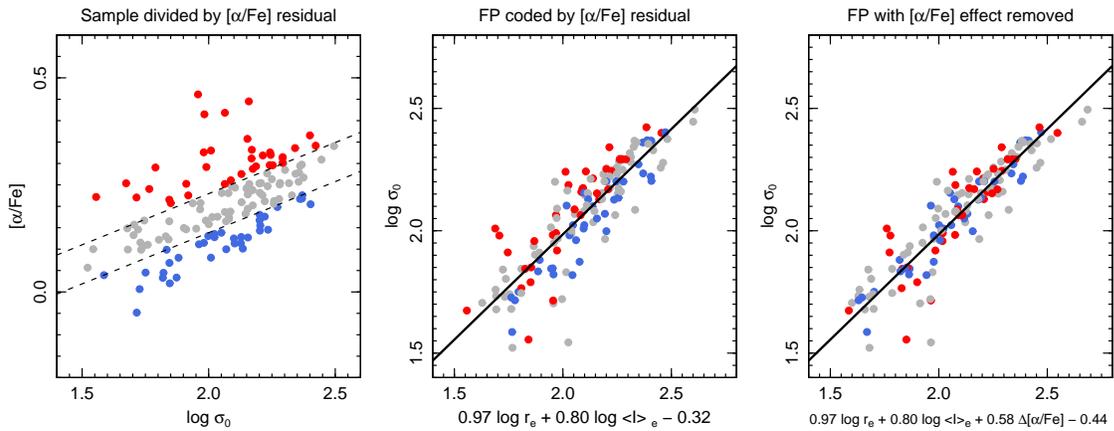}\\
  \caption{Illustration of the $\alpha$/Fe dependence of the FP
    residuals. In the left-hand panel we define three subsamples using
    residuals from the $\alpha$/Fe-log$\sigma$ relation. The 50$\%$ of
    galaxies closest to the relation are plotted in grey, while the
    25$\%$ with $\alpha$/Fe higher (lower) than expected for their $\sigma$
    are plotted in red (blue). The central panel shows the edge-on
    projection of the FP colour-coded according to this scheme. It is
    readily seen that high-$\Delta$[$\alpha$/Fe] galaxies are offset
    from the mean FP in a direction which may be interpreted as
    smaller effective radius, lower surface brightness, or higher
    velocity dispersion. The low-$\Delta$ $\alpha$/Fe galaxies are
    offset from the mean FP in the opposite direction. In the
    right-hand panel, we show the FP after correcting for the $\alpha$/Fe
    correlation. A modest reduction in total scatter is seen, despite
    the introduction of uncertainty due to $\alpha$/Fe measurement errors.}
\label{split}
\end{figure*}

While it seems that none of the additional spp terms reduce the
overall scatter significantly, the relative impact of the three spp
terms becomes clearer when considering the intrinsic scatter
(i.e. after accounting for the measurement uncertainties).  Both the
relation including the effect of age and metallicity have an intrinsic
scatter of ${\sim}0.06$, only marginally lower than the value without
the spp term, while in the case of $\alpha$/Fe the intrinsic scatter
is reduced to 0.049. This is comparable to the rms contribution from
the $\alpha$/Fe term (0.047), i.e. $\alpha$/Fe contributes around half
of the intrinsic scatter, and indicates that the distribution of
galaxies around the FP are tightly related to the enrichment, and
hence to the timescale of star-formation. The $\alpha$/Fe dependence
of the FP scatter is illustrated further in Figure~\ref{split} which
demonstrates also the reduction in total scatter despite the
introduction of measurement errors in $\alpha$/Fe.  We split the
Shapley sample into three groups (shown as red, grey and blue points
in Figure~\ref{split}) according to the position of each galaxy with
respect to the $\alpha$/Fe-log\,$\sigma_{0}$ relation (see left
panel): the 50$\%$ of galaxies closest to the relation are plotted in
grey, while the 25$\%$ with $\alpha$/Fe higher (lower) than expected
for their $\sigma$ are plotted in red (blue). A modest reduction in
total scatter is seen, despite the introduction of uncertainty due to
$\alpha$/Fe measurement errors.

Furthermore the total FP scatter is reduced by
including the extra terms, in spite of the fact that the SSP
parameters are subject to their own substantial uncertainties. This
implies there must be a still greater reduction in the {\it intrinsic}
FP scatter.

\section{The origin of the intrinsic scatter}

JFK96 pointed out that the dispersion around the FP relation is not
completely due to the measurement errors but has an intrinsic scatter
whose nature is not yet understood. The existence of this intrinsic
scatter was interpreted as due to another ``fundamental'' parameter
characterising the family of early-type galaxies. In many of the
previous works on the FP (see Section 1), a strong effort has been
made to find correlations between the FP residuals and different
line-indices considered to be representative of a particular stellar
population parameter (for example H$\beta$ for age, Mg$_2$ for
metallicity etc).

In Section 5 we showed that the FP residuals are strongly
correlated with estimated stellar population parameters especially
with the $\alpha$-abundance ratio ($\alpha$/Fe).  Given that age
and $\alpha$/Fe are known to correlate strongly for galaxies of a
given mass, while age and metallicity anti-correlate (e.g. Proctor
\& Sansom 2002; Smith et al. 2007b), it is not clear which of the
three parameters drives the correlations. However, the fact that
the strongest correlation is observed for the $\alpha$/Fe suggests
that this parameter is playing the major role, while age and
metallicity trends just reflect their mutual correlations. The
trends observed for the metallicity indices also suggest that the
$\alpha$/Fe is driving the correlations. In fact, the positive
correlations between the Z/H dependent indices Fe4383 and Fe5015
combined with the negative correlation for Mgb, is consistent with
an anti-correlation for the abundance ratio $\alpha$/Fe. Moreover,
we found that about half of the FP intrinsic variance is due to
variations in the stellar populations as described by the
$\alpha$-enhancement trend, while the other stellar population
parameters seem to contribute rather less to the intrinsic
scatter.

It has long been realised (Tinsley 1980) that in the study of galaxy
formation and evolution, a major role can be played by the analysis of
the abundance ratios, due to them being relatively model independent
and primarily affected by stellar nucleosynthesis and the initial mass
function (Matteucci 1996). In particular, the ratio between the
so-called $\alpha$ elements (synthesized in type II supernovae) and
iron (mainly from delayed type Ia events) is widely accepted to be
affected by the characteristic timescale of star-formation, the
$\alpha$ elements and iron having different production timescales
(Greggio \& Renzini 1983). According to the simplest and widely
accepted scenario, a galaxy with a high value of $\alpha$/Fe has
experienced many type II SNe events, but almost no SNe of type Ia
during the major epoch of star-formation: this constrains the
timescale for this epoch of stellar formation to be shorter than
3$\times$10$^{8}$ years.

Our results suggest that the galaxies which formed their stars over a
shorter duration (high $\alpha$/Fe) are also those which are more
compact. Such a pattern may be expected in a dissipational merger
inducing a nuclear sturburst.  Vazdekis, Trujillo and Yamada (2004),
studing 21 early-type galaxies already claimed the existence of a
correlation between Sersic index and Mg/Fe ratio (i.e. between galaxy
structure and stellar population), even if they did not correct for
correlations with $\sigma$. They interpreted their results as more
massive galaxies having their star formation quenched on shorter
time-scales. Although, in agreement with our finding, their analysis
is based on to a smaller sample spanning a narrower range of velocity
disperisons and luminosities with compared to our Shapley sample
making any comparisons difficult.

\citet{hopkins} make specific predictions for the effects of varying
the dissipational fraction on the remnant spheroids of a given mass,
resulting in correlations between the structural and stellar
population parameters. By considering identical progenitor disks (at
$t=0$) with initial gas fractions $f_{gas}=1$ following an exponential
star-formation history with time-scale $\tau$, we have that the gas
fraction at the time of the merger $t_{m}$ (and hence dissipational
fraction in the merger-induced starburst) will scale as
$f_{gas}=\exp(-t_{m}/\tau)$. If the remaining gas is then consumed in
the central star-burst, producing a passively-evolving spheroid
remnant, then the dissipational fraction will increase for earlier
mergers, and hence mean stellar age. \citet{hopkins} also show that
ellipticals formed through mergers with higher dissipation fractions
should be more $\alpha$-enriched. As merger remnants involving more
dissipation, like those between more gas-rich disks, are expected to
yield larger mass fractions formed in nuclear starbursts, which reduce
significantly the effective radii of the remnant, we should expect
both mean stellar age and $\alpha$/Fe to anti-correlate with the
residuals in the $r_{\rm e}$ direction.

We observe clear correlations between the FP residuals and age and
$\alpha$/Fe, with those galaxies of a given mass with effective radii
smaller than predicted by the FP to have stellar populations
systematically older and with higher abundances than average, fully
consistent with the predictions of \citet{hopkins}.

\section{Summary and Conclusions}

We have derived the FP of a sample of 141 early-type $R{<}18$ galaxies
in the Shapley supercluster at $z{=}0.049$.  Velocity dispersions and
stellar population parameters were derived from the spectroscopic data
of Smith et al.  (2007), while R-band photometry is from the Shapley
Optical Survey \citep{mercurio}. The final sample extends down to
M$_{R}^{*}$+3 in magnitude and 50 km s$^{-1}$ in $\sigma_{0}$.  Using
the software 2DPHOT (La Barbera et al. 2008b) we derived for each
galaxy the structural parameters $r_{\rm e}$,
${\langle}\mu{\rangle}_{\rm e}$, $n$, and $m_{tot}$ by fitting a 2D
PSF-convolved Sersic model. The morphological classification was
performed by eye and checked with those of \cite{thomas06} for a
subsample of 54 galaxies finding complete agreement.

Adopting a weighted fit (see Eq.~\ref{fp_weighted}) we derived the FP:
$r_{\rm e}$\,${\propto}$\,$\sigma^{1.06} {\langle}I{\rangle}_{\rm
  e}^{-0.82}$ for the 141 early-type R$<$18 galaxies. The low value of
$\alpha$ can be related to the extension of our sample towards very
low values of velocity dispersion and and also to the method used to
determine $\sigma$.  Considering the subsample of $ \sigma_{0}>100$ km
s$^{-1}$ galaxies the FP turns out to follow the relation $r_{\rm
  e}$\,${\propto}$\,$\sigma^{1.35} {\langle}I{\rangle}_{\rm
  e}^{-0.81}$ which is consistent with the Coma FP obtained by JFK96.
Observing the significant change in the $\alpha$ value between the
total and the high-$\sigma_{0}$ samples, we investigate the possible
curvature of the FP. Applying the same cuts to a simulated sample
following a linear FP relation, we ascertain that the observed
curvature can be explained by selection effects. Departure from a
linear trend of the FP relation has been claimed by D'Onofrio et
al. (2008), and Desroches et al. (2001) found that the faint-end
luminosity cut influences the coefficient $\alpha$ of the FP, in
agreement with Nigoche-Netro (2008). Similar results were found by
Hyde $\&$ Bernardi (2008) studying the FP residuals along the
plane. However, these works pointed out that conclusive evidence of
the FP curvature needs either more robust statistics or higher
accuracy in the velocity dispersion estimates for low mass systems.

The most important result of this work is our demonstration that the
FP residuals are correlated with stellar population characteristics
(line- strength indices and derived SSP parameters). In particular, FP
residuals are anti-correlated both with the $\alpha$-element abundance
ratio, $\alpha$/Fe, and with galaxy age resulting in trends whereby
galaxies more compact than expected from the FP relation have stellar
pupulations systematically older and with higher abundances than
average. Previous studies have reported correlations of the FP
residuals with stellar age (e.g. Forbes et al. 1998; Reda et
al. 2005). Although our FP residual do show a correlation with age, we
recover a much stronger signal for $\alpha$/Fe than for age,
suggesting that this is the more fundamental dependence. Indeed, a
multiple regression analysis suggests there is no age correlation at
fixed $\alpha$/Fe, but a very strong $\alpha$/Fe correlation at fixed
age. The correlation between $\alpha$/Fe and FP residuals does not
indicate a direct {\it causal} effect, since varying $\alpha$/Fe at
fixed age and Z/H has little effect on the stellar
mass-to-light. Instead, the correlation suggests that the structural
properties and the star-formation history are both dependent on some
unobserved aspect of the galaxy assembly process. At face value, our
results are broadly consistent with recent galaxy merger simulations,
which predict a sequence of formation mechanisms governed by the
varying importance of dissipation (Hopkins et al. 2008).  In this
scenario, mergers with a higher initial gas-fraction trigger more
centrally-concentrated starbursts, and higher $\alpha$ abundances due
to the short duration of star-formation in the burst.

\section*{Acknowledgments}

AG gratefully acknowledges the hospitality of the University of
Birmingham during her stays there, where some of the work was
performed. CPH acknowledges financial support from STFC. RJS is
supported by STFC rolling grant PP/C501568/1 ``Extragalactic Astronomy
and Cosmology at Durham 2005--2010".

This work was carried out in the framework of the collaboration of
to the FP7-PEOPLE-IRSES-2008 project ACCESS ``A Complete CEnsus of
Star-formation and nuclear activity in the Shapley supercluster".

We thank the anonymous referee for his/her comments which helped
to improve this work.

\label{lastpage}
\end{document}